\documentclass[10pt,twocolumn,letterpaper]{article}

\usepackage[pagenumbers]{cvpr} 

\usepackage{amsfonts}
\usepackage{amsmath}
\usepackage{booktabs}
\usepackage{multirow}

\usepackage{xcolor}

\usepackage{algorithm}
\usepackage{algpseudocode}
\usepackage{url}
\usepackage{graphicx}
\usepackage{booktabs}
\usepackage{multirow}
\usepackage{wrapfig}
\usepackage{xcolor}

\definecolor{cvprblue}{rgb}{0.21,0.49,0.74}
\usepackage[pagebackref,breaklinks,colorlinks,allcolors=cvprblue]{hyperref}

\def\confName{CVPR}
\def\confYear{2027}

\title{PSM-WM: Predictively Sufficient Multimodal Social World Modeling}

\author{
Chunlei Meng$^{1}$,
Zhenyu Yu$^{1}$,
Pengbin Feng$^{2}$,
Jacqueline J. Pang$^{3}$\\
Jiabin Luo$^{4}$,
Chih-Ting Liao$^{5}$,
Zhaolu Kang$^{4}$,
Chun Ouyang$^{1,*}$,
Zhongxue Gan$^{1}$
\\
\vspace{0.2em}
{\tt\small clmeng23@m.fudan.edu.cn}
\\
\vspace{0.3em}
$^1$ Fudan University
$^2$ University of Southern California\\
$^3$ Cornell University
$^4$ Peking University
$^5$ University of New South Wales
}

\begin{document}
\maketitle
\begin{abstract}
Learning latent states that support future evolution remains an open problem
for multimodal interactive systems. Existing approaches typically optimize
representations for current semantic understanding or direct prediction,
without explicitly defining what information a latent state should preserve
to remain predictive of future dynamics. In this work, we investigate this
problem through the principle of predictive sufficiency, which defines a
predictive state as a latent variable that retains future-relevant
information from historical multimodal observations while reducing redundant
dependence on the observed history. Based on this principle, we propose Predictively Sufficient Multimodal Social World Modeling (PSM-WM), a framework that jointly learns predictive latent states and their stochastic temporal evolution through probabilistic state inference, event-conditioned transitions, and multi-step consistency optimization. Beyond forecasting performance, we introduce state-centric evaluations to measure whether learned representations satisfy predictive state properties, including predictive closure, residual history gain, and future semantic alignment. Experiments on Hi-EF, IEMOCAP, MELD, and Ego4D show that PSM-WM improves future multimodal interaction modeling and learns latent states with stronger predictive properties and robustness. These results indicate that predictive sufficiency provides an effective principle for constructing future-oriented
multimodal states and highlight the importance of state construction in
multimodal world modeling. 
\end{abstract}    
\begin{figure}
    \centering
    \includegraphics[width=1\linewidth]{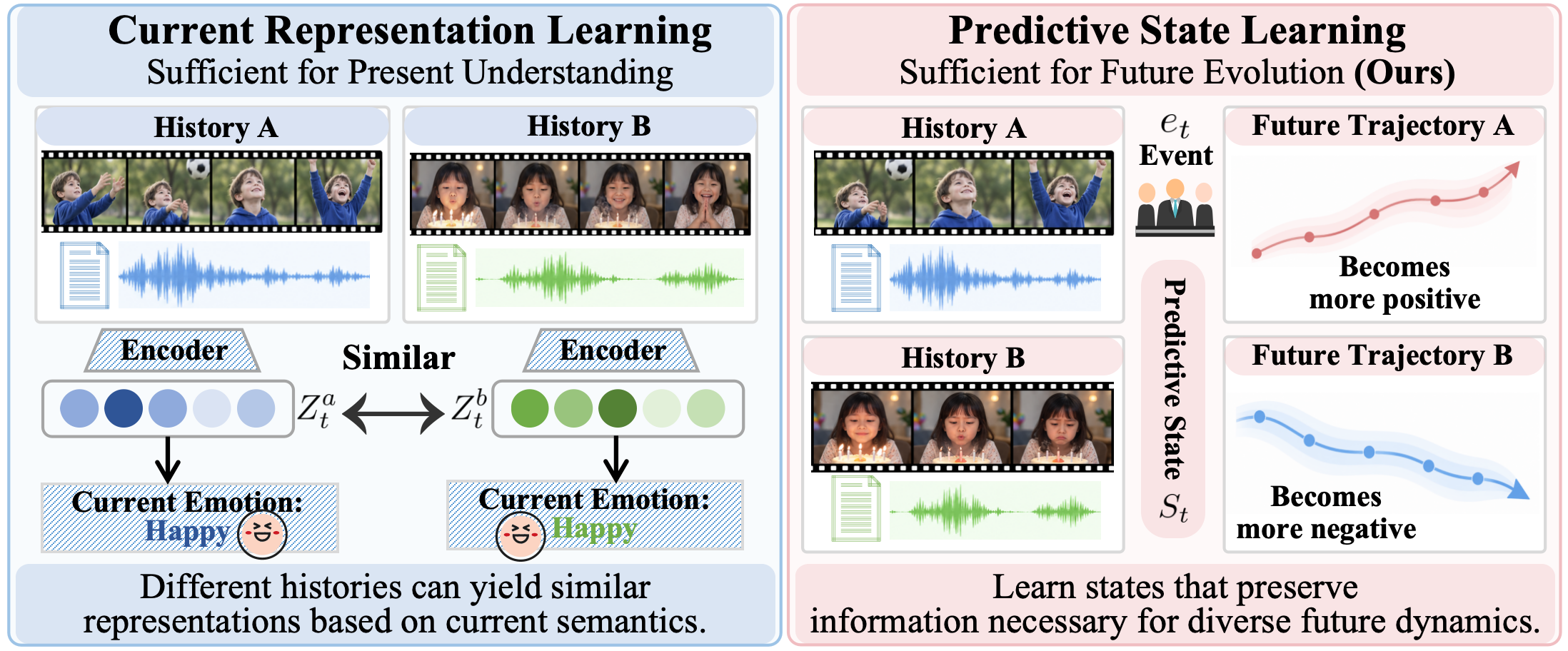}
    \caption{Motivation of predictive state learning. Identical current semantics may correspond to different future dynamics. Predictive state learning aims to construct latent states that preserve information necessary for future evolution.}
    \label{fig:motivation}
\end{figure}

\section{Introduction}
\label{sec:intro}

Building intelligent systems capable of interacting with dynamic environments
requires internal states that not only describe the present but also support
future reasoning and evolution. Recent advances in world models have
demonstrated the importance of learning latent dynamics for predicting future
states and enabling long-horizon decision making~\cite{DreamerV3}. Meanwhile, predictive representation learning has shown
that representations optimized for future information can provide more
generalizable abstractions beyond conventional discriminative objectives
~\cite{lecun2022path}. These studies collectively
suggest that future-oriented representations are essential for intelligent
systems. However, a fundamental question remains open: what information
should a latent state preserve such that future evolution can be predicted
from the state itself?

Multimodal social interaction provides a challenging scenario for studying
this question. Recent multimodal representation learning methods have achieved
significant progress in modeling human behaviors by exploiting cross-modal
interaction, modality-specific information, and semantic alignment~\cite{DLF,misa,GCL,MRCF}.
Transformer-based approaches learn flexible interactions among heterogeneous
modalities~\cite{MULT}, while representation disentanglement
methods such as MISA~\cite{misa} and subsequent multimodal
learning frameworks~\cite{CLCR,TSD,EMOE} further improve the separation and integration of complementary information. However, these approaches mainly evaluate representations according to their ability to explain current observations or predict current semantic states. Such representations may preserve discriminative information for the present while discarding historical factors that determine future interaction
dynamics. As illustrated in Fig.~\ref{fig:motivation}, identical current
semantics may correspond to different future dynamics.

Beyond representation learning, future forecasting methods explicitly model
temporal evolution by predicting future outcomes from historical observations.
Although these methods demonstrate that historical context contains useful
future information, they typically formulate forecasting as a direct mapping $H_t \rightarrow F_{t+k},$ without explicitly characterizing whether the learned latent representation itself contains the information required for future evolution. Similarly, predictive representation learning methods optimize future consistency between representations~\cite{lecun2022path}, but generally focus on
learning predictive embeddings rather than defining the principle of a
predictive state. Therefore, accurate future prediction does not necessarily
imply that a model has discovered a reusable latent state. World models provide another perspective by learning latent dynamics that
describe how states evolve over time. Existing latent dynamics models,
including PlaNet and Dreamer-style architectures, learn transition mechanisms
over latent variables for long-horizon prediction~\cite{RSSM,DreamerV3}. Recent social world models further study how social beliefs
evolve in response to events~\cite{SWM}. However, most
world models assume that meaningful latent states are already available or
focus primarily on transition modeling. Consequently, an important challenge
remains unresolved: How should a latent state be constructed such that it preserves exactly the information necessary for future evolution? 

We investigate this problem from the perspective of predictive sufficiency.
Inspired by the principle of sufficient statistics, we define a predictive
state as a latent variable that preserves future-relevant information from
historical multimodal observations while reducing the additional predictive dependence on the observed history. Formally, an ideal predictive state $S_t$ should
satisfy $p(F_{t+1:t+k}|H_t,E_t)\approx p(F_{t+1:t+k}|S_t,E_t),$ which indicates that once the predictive state is known, the remaining historical observations provide limited additional information about future evolution. Equivalently, this objective can be interpreted as minimizing $I(F_{t+1:t+k};H_t|S_t,E_t).$ This perspective shifts multimodal representation learning from constructing states that explain current observations toward learning states that preserve future evolution dynamics.

\begin{figure*}
    \centering
    \includegraphics[width=0.9\linewidth]{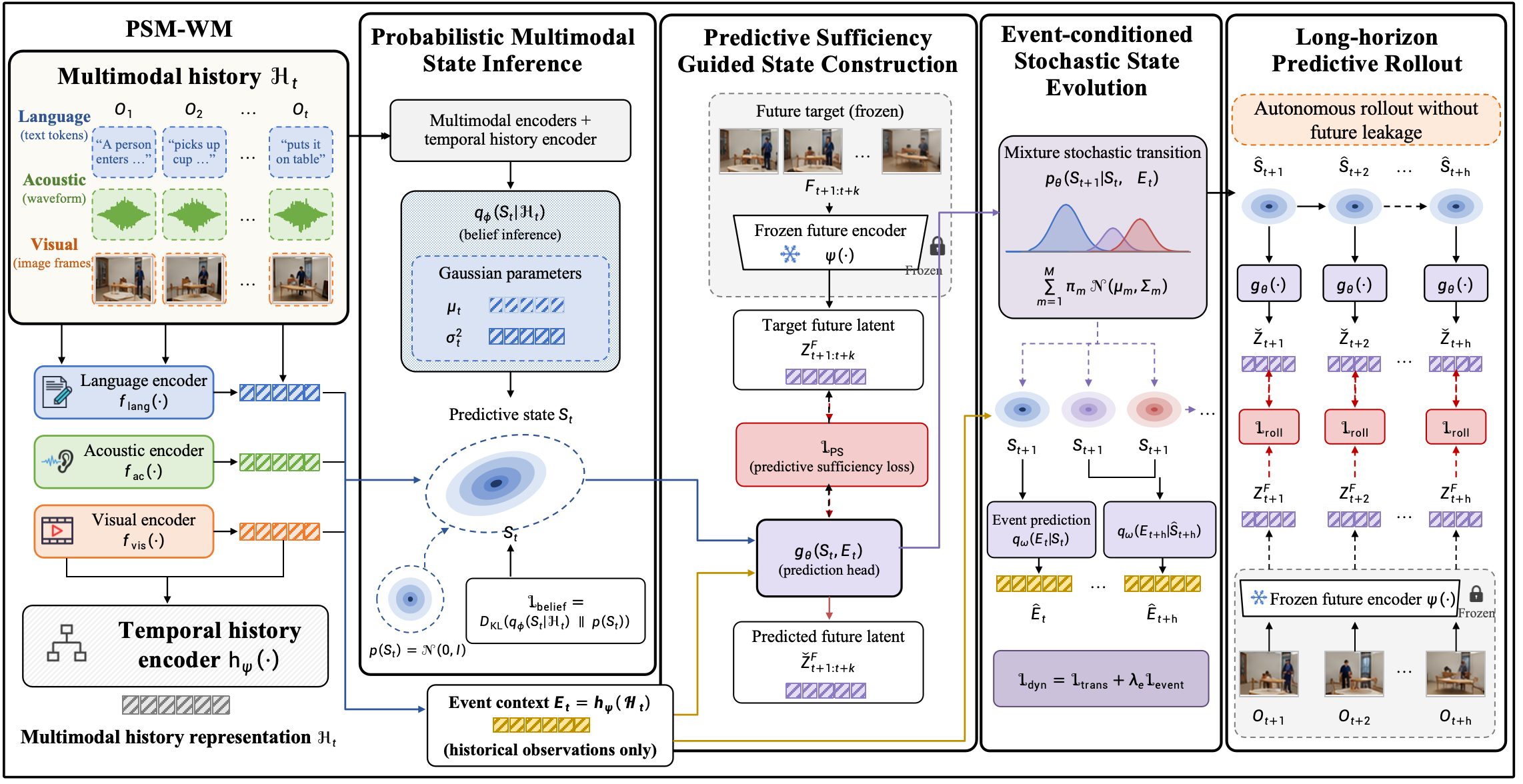}
    \caption{The framework learns predictive multimodal states through probabilistic belief state inference, predictive sufficiency guided state construction, and event-conditioned stochastic transition modeling, enabling long-horizon prediction of multimodal interaction dynamics.}
    \label{fig:PSM-WM}
\end{figure*}

Based on this principle, we propose Predictively Sufficient Multimodal Social World Modeling (PSM-WM), a framework for learning future-oriented multimodal latent states in social interactions (Fig.~\ref{fig:PSM-WM}). We introduce predictive sufficiency as a state learning criterion, which separates two fundamental aspects of world models: what information should be preserved in a latent state and how such a state evolves over time. PSM-WM consists of three complementary
components: probabilistic belief state inference for modeling uncertainty from partially observed multimodal histories, predictive sufficiency guided state construction for preserving future-relevant information, and event-conditioned stochastic transition modeling for capturing latent state evolution. Furthermore, we introduce state-centric evaluation protocols that
go beyond conventional forecasting accuracy and examine whether learned states retain future predictive information. Experiments on Hi-EF, IEMOCAP, MELD, and Ego4D show that optimizing latent states according to predictive sufficiency improves future multimodal interaction modeling. Our contributions are summarized as follows:
\begin{itemize}
\item We identify predictive sufficiency as a fundamental criterion for
constructing future-oriented multimodal states.

\item We develop PSM-WM, a framework that operationalizes this criterion through
probabilistic state inference and stochastic latent evolution modeling.

\item We establish state-centric evaluations and empirical analyses to examine
whether learned representations preserve future-relevant information beyond
direct forecasting.
\end{itemize}

\section{Related Work}

\textbf{Multimodal Representation Learning.}
Existing multimodal methods primarily focus on learning discriminative
representations through cross-modal interaction, modality-specific
modeling, and uncertainty-aware fusion~\cite{TSDA,Aoept,AHRA}.
Recent approaches, such as EMART~\cite{EMART} and
HyperEF~\cite{HyperEF}, further improve representation quality
through emotion-structure-aware learning and missing-modality
recovery. However, these methods mainly optimize current-state
understanding, leaving the preservation of future-relevant information
largely unexplored.

\textbf{Multimodal Future Forecasting.}
Recent work extends multimodal understanding to future interaction
prediction~\cite{DBR,Hi-EF}, showing that historical
observations contain useful signals for forecasting future states.
Nevertheless, such approaches mainly formulate forecasting as
a direct mapping from history to future outcomes, without
explicitly characterizing the latent state required for future evolution.

\textbf{World Models and Predictive State Learning.}
World models, including RSSM~\cite{RSSM}, DreamerV3~\cite{DreamerV3},
and recent stochastic dynamics approaches~\cite{LPWM,HAUWM},
learn latent representations and model their temporal evolution.
However, their primary focus is on latent transition modeling rather
than explicitly establishing the predictive sufficiency of the
underlying state. In contrast, we investigate state construction
through the predictive sufficiency principle,
formulated as
$P(F\mid H,E)\approx P(F\mid S,E)$,
and develop a multimodal world model that jointly learns probabilistic
state inference and event-conditioned stochastic transitions.

A more comprehensive discussion of related methods and their
differences is provided in Appendix~A.

\section{Method}

\subsection{Problem Formulation and Predictive State Principle}
\label{sec:formulation}

We consider multimodal social interaction sequences consisting of
language, acoustic, and visual observations. At time step $t$, the observed
multimodal history is denoted as $H_t=\{O_1,O_2,\cdots,O_t\},$ where each observation contains multiple modalities $O_t=\{X_t^l,X_t^a,X_t^v\}.$ Given the historical context $H_t$, the goal is to learn a latent state $S_t$ that supports future interaction evolution. Unlike conventional representation learning, which optimizes states for current prediction objectives, we define a predictive state by the information required for future dynamics.

\textbf{Conditional Predictive State Principle.}
A meaningful latent state should preserve information required for future interaction evolution while reducing the additional predictive dependence on historical observations. Instead of assuming that the latent state alone determines future dynamics,
we define predictive states together with an observable event context. Specifically, a predictive state $S_t$ should satisfy, 
\begin{equation}
 p(F_{t+1:t+k}|H_t,E_t) \approx p(F_{t+1:t+k}|S_t,E_t),   
\end{equation}
where $F_{t+1:t+k}$ denotes the future interaction trajectory within a predefined prediction horizon. That is, once the state and event context are known, the remaining history adds little information for the evaluated prediction task.

\subsection{Probabilistic Multimodal State Inference}

The latent social state cannot be directly observed from multimodal
interaction signals. Instead, it must be inferred from heterogeneous
observations that may contain ambiguity and incomplete evidence. Therefore, we represent the latent state as a probabilistic belief distribution $q_\phi(S_t|H_t)
=\mathcal{N}(\mu_t,\sigma_t^2),$ where $\mu_t$ represents the estimated latent state and $\sigma_t^2$ captures uncertainty induced by incomplete observations. To regularize the latent state space and prevent arbitrary uncertainty estimation, we introduce a prior distribution $p(S_t)=\mathcal{N}(0,I),$ and optimize the belief regularization objective:
\begin{equation}
    L_{belief}=D_{KL}(q_\phi(S_t|H_t)||p(S_t)).
\end{equation}

The inferred belief state is the basis for the subsequent state construction and transition learning.

\subsection{Predictive Sufficiency Guided State Construction}

The central problem of predictive state learning is determining what
information should be preserved in the latent state; we address it through
predictive sufficiency.

Given the inferred belief state $S_t$, an ideal predictive state should
preserve all future-relevant information contained in the historical
context. Following the predictive sufficiency principle introduced in
Sec.~\ref{sec:formulation}, this requirement can be expressed as:
\begin{equation}
    p(F_{t+1:t+k}|H_t,E_t)\approx p(F_{t+1:t+k}|S_t,E_t).
\end{equation}
Equivalently, the residual predictive information from the historical context after observing the latent state should be minimized $I(F_{t+1:t+k};H_t|S_t,E_t)\rightarrow0.$ Directly optimizing this objective is challenging because the complete future distribution is unknown and high-dimensional. Therefore, we derive a practical approximation by maximizing the predictive information retained in the latent state. Specifically, we introduce a future representation encoder $\psi(\cdot)$ that extracts future semantic states $Z^F_{t+1:t+k}=\psi(F_{t+1:t+k}),$ and learn a predictive mapping from the current latent state $\hat Z^F_{t+1:t+k}=g_\theta(S_t,E_t).$ The predictive sufficiency objective is approximated by minimizing the distance between predicted and target future representations:
\begin{equation}
    \mathcal{L}_{pred}=d(\hat Z^F_{t+1:t+k},Z^F_{t+1:t+k}).
\end{equation}
Importantly, $\mathcal{L}_{pred}$ serves as an empirical surrogate for predictive information preservation. The future encoder $\psi$ is pretrained and frozen during optimization, so $Z^F$ provides a fixed predictive target rather than an adaptive representation.

\subsection{Event-conditioned Stochastic State Evolution}

A predictive state should not only preserve future-relevant
information, but also support modeling how latent interaction states
evolve over time. However, social interactions are inherently
stochastic and may contain multiple plausible future trajectories.
Therefore, deterministic latent transitions are insufficient for
capturing the uncertainty and diversity of future evolution. Given the predictive state $S_t$, we first extract the current interaction event from available observations $E_t=h_\psi(H_t),$ where $h_\psi(\cdot)$ only receives historical observations available at time $t$. To model the stochastic nature of social evolution, we formulate the
latent transition as a mixture distribution:
\begin{equation}
\begin{aligned}
&p_\theta(S_{t+1}|S_t,E_t)
=\sum_{m=1}^{M}\pi_m(S_t,E_t)\\
&\qquad{}\times\mathcal{N}\!\left(
\mu_m(S_t,E_t),\sigma_m^2(S_t,E_t)
\right),
\end{aligned}
\end{equation}
where $\pi_m$ represents the probability of each possible evolution
mode. This formulation enables the world model to maintain multiple
future hypotheses instead of collapsing uncertain trajectories into a single averaged prediction.

To enable autonomous future rollout, we further learn an event predictor that estimates interaction conditions from latent states.
Specifically, the event inference network is optimized by $L_{event}=-\log q_\omega(E_t|S_t),$ where the event label $E_t$ is obtained from the historical context encoder $h_\psi(H_t)$. This objective
encourages the latent state to retain the information needed to predict interaction conditions.

During inference, future interaction events are unavailable.
Therefore, instead of directly accessing future observations, we
infer future event distributions from predicted latent states $q_\omega(E_{t+h}|S_{t+h}),$ which enables autonomous latent rollout without future information leakage. The transition process during rollout is therefore, $\hat{E}_{t+h}
\sim q_\omega(E|\hat{S}_{t+h}),$ followed by, $\hat{S}_{t+h+1}
\sim p_\theta(S_{t+h+1}|\hat{S}_{t+h},\hat{E}_{t+h}).$ The transition and event prediction objectives are jointly optimized as:
\begin{equation}
L_{dyn}
=
-\log p_\theta(S_{t+1}|S_t,E_t)
-\lambda_e
\log q_\omega(E_t|S_t).
\end{equation}

\subsection{Overall Optimization Objective}

The objective of PSM-WM is to learn latent states that preserve
future-relevant information and support consistent future evolution.
Accordingly, the optimization objective consists of five complementary
terms corresponding to different properties of predictive states.

First, the belief regularization objective constrains the probabilistic
state inference:
\begin{equation}
\mathcal{L}_{belief}
=
D_{KL}
(
q_\phi(S_t|H_t)
||
p(S_t)
). 
\end{equation}
\begin{equation}
\mathcal{L}_{PS}
=
d(
g_\theta(S_t,E_t),
Z^F_{t+1:t+k}
). 
\end{equation}
This objective regularizes the latent state distribution and enables
uncertainty-aware state inference. Second, predictive sufficiency is approximated by future representation prediction $\mathcal{L}_{PS}$. Third, the transition objective learns the evolution mechanism between
predictive states:

\begin{equation}
\mathcal{L}_{trans}
=
-\log
p_\theta(S_{t+1}|S_t,E_t).
\end{equation}
\begin{equation}
\mathcal{L}_{roll}
=
\sum_{h=1}^{H}
\gamma^{h-1}
d(
\hat Z_{t+h},
Z^F_{t+h}
).
\end{equation}

Finally, to improve long-horizon consistency, we recursively apply the
transition model and optimize future latent alignment $\mathcal{L}_{roll}$. The final training objective is:
\begin{equation}
L=
L_{belief}
+\lambda_1L_{PS}
+\lambda_2L_{trans}
+\lambda_3L_{event}
+\lambda_4L_{roll}.
\end{equation}
Unlike
direct forecasting objectives that only optimize future prediction accuracy, PSM-WM explicitly constrains what information should be retained in the latent state and how such states should evolve over time.

\section{Experiments}

\subsection{Experimental Goals}
Our experiments aim to evaluate whether predictive sufficiency provides
an effective principle for constructing future-oriented multimodal states.
Specifically, we investigate four questions: 

\textbf{RQ1:} Can predictive state construction improve future multimodal evolution modeling across different environments? 

\textbf{RQ2:} Do the learned latent states preserve future-relevant information and satisfy the properties of predictive sufficiency? 

\textbf{RQ3:} How do state construction and stochastic transition modeling contribute to stable long-horizon latent evolution? 

\textbf{RQ4:} Can predictive states remain reliable under incomplete observations and uncertain future dynamics? 


\begin{figure*}[htbp]
    \centering
    \includegraphics[width=1\linewidth]{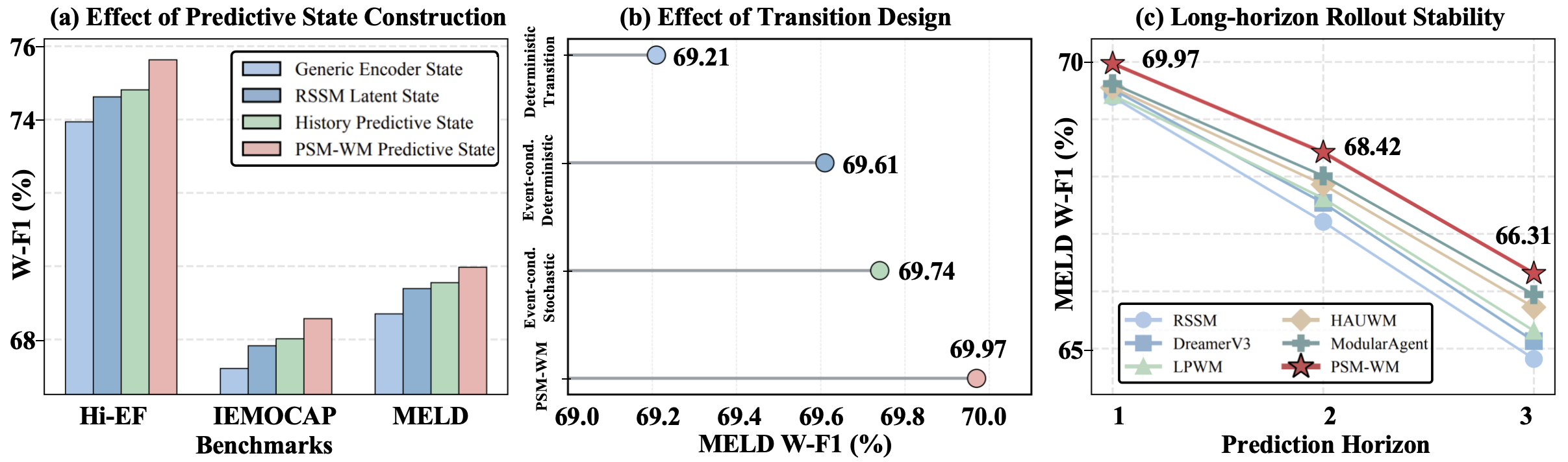}
   \caption{Predictive state evolution analysis. (a) Predictive states improve latent representations under fixed transitions. (b) Event-conditioned stochastic dynamics enhance state evolution. (c) Predictive states enable stable long-horizon rollout.}
    \label{fig:PSE-Analysis}
\end{figure*}

\subsection{Experimental Setup}
\label{sec:setup}

\textbf{Datasets.}
We evaluate PSM-WM on four multimodal temporal benchmarks:
Hi-EF~\cite{Hi-EF}, IEMOCAP~\cite{IEMOCAP}, MELD~\cite{meld},
and Ego4D~\cite{Ego4D}. Hi-EF, IEMOCAP, and MELD provide temporally evolving multimodal interaction scenarios with aligned language, acoustic, and visual signals, enabling the evaluation of predictive state learning in social environments. To assess generalization beyond social interactions, we additionally evaluate PSM-WM on Ego4D, which captures long-duration human-environment interactions with visual and language observations. \textbf{Evaluation Metrics.}
We evaluate models from three complementary perspectives. First, future prediction performance is measured using weighted F1 score
(W-F1) on social interaction benchmarks and future activity prediction accuracy on Ego4D. Second, predictive state quality is evaluated using state-centric metrics: Predictive Closure Gap (PCG), which measures the performance difference between full-history prediction and state-only prediction; Future Latent Retrieval (FLR), which evaluates whether the learned state contains information predictive of future representations; and Residual History Gain (RHG), which measures the additional predictive benefit obtained by providing the original history after observing the latent state. Third, state evolution quality is evaluated using long-horizon rollout performance. \textbf{Implementation Details.} All methods use identical multimodal encoders for fair comparison. Results are averaged over multiple random seeds, and mean and standard deviation are reported.

\begin{table}[t]
\centering
\setlength{\tabcolsep}{5pt} 
\renewcommand{\arraystretch}{1.05}
\caption{Future interaction prediction performance. Higher W-F1 indicates better future modeling capability.}
\begin{tabular}{lccc}
\toprule
Method & Hi-EF & IEMOCAP & MELD\\
\midrule
\multicolumn{4}{c}{\textbf{\textit{Multimodal Representation Learning}}}
\\
MISA~\cite{misa}&71.24&65.31&66.52\\
EMOE~\cite{EMOE}&72.18&66.02&67.14\\
TSD~\cite{TSD}&72.83&66.71&68.21\\
EMART~\cite{EMART}&73.12&67.03&68.64\\
HyperEF~\cite{HyperEF}&73.84&67.45&68.92\\
\midrule
\multicolumn{4}{c}{\textbf{\textit{Predictive Representation Learning}}}
\\
CPC~\cite{CPC}&72.96&67.18&67.86\\
BYOL~\cite{BYOL}&73.21&67.42&68.14\\
V-JEPA~\cite{V-JEPA2}&74.10&67.85&68.95\\
\midrule
\multicolumn{4}{c}{\textbf{\textit{Predictive State and World Models}}}
\\
RSSM~\cite{RSSM}&74.62&67.83&69.39\\
DreamerV3~\cite{DreamerV3}&74.80&68.02&69.52\\
LPWM~\cite{LPWM}&74.65&68.05&69.42\\
HAUWM~\cite{HAUWM}&74.75&68.10&69.55\\
ModularAgent~\cite{ModularAgent}&74.91&68.15&69.62\\
\midrule
\textbf{PSM-WM (Ours)}&\textbf{75.76}&\textbf{68.61}&\textbf{69.97}\\
\bottomrule
\end{tabular}
\label{tab:main_results}
\end{table}

\subsection{RQ1: Predictive States Improve Future Multimodal Evolution Modeling}

We first investigate whether constructing latent states according to
predictive sufficiency improves future multimodal evolution modeling.
To this end, we compare PSM-WM with three categories of approaches:
multimodal representation learning methods (MISA~\cite{misa}, EMOE~\cite{EMOE}, TSD~\cite{TSD}, EMART~\cite{EMART}, and HyperEF~\cite{HyperEF}), predictive representation learning methods (CPC~\cite{CPC}, BYOL~\cite{BYOL}, and V-JEPA~\cite{V-JEPA2}), and latent dynamics models (RSSM~\cite{RSSM}, DreamerV3~\cite{DreamerV3}, LPWM~\cite{LPWM}, HAUWM~\cite{HAUWM}, and ModularAgent~\cite{ModularAgent}). As shown in Table~\ref{tab:main_results}, PSM-WM achieves the best
performance across Hi-EF, IEMOCAP, and MELD. Compared with multimodal
representation learning methods, the improvements indicate that
representations optimized for current semantic understanding are less
effective for modeling future evolution. Compared with predictive
representation learning methods, PSM-WM further benefits from explicitly constructing latent states according to future-relevant information preservation. Moreover, PSM-WM consistently outperforms latent dynamics baselines, achieving improvements of 0.85, 0.46, and 0.35 percentage points in W-F1 over the strongest world model baseline, ModularAgent, on Hi-EF, IEMOCAP, and MELD, respectively. These results demonstrate that effective future multimodal modeling depends not only on learning transition dynamics, but also on constructing latent states that preserve information required for future evolution.

\subsection{RQ2: Analysis of Predictive State Properties}

Future prediction performance alone does not reveal whether a model learns a predictive state, since accurate forecasting can still rely on implicit access to historical observations. Therefore, we evaluate whether the latent states learned by PSM-WM preserve future-relevant information under a controlled state-level probe protocol. Specifically, we examine the three state-centric metrics defined in Sec.~\ref{sec:setup} (PCG, RHG, and FLR). All evaluations use frozen representations and identical prediction probes, ensuring that the measured differences originate from the learned states rather than predictor capacity.

As shown in Table~\ref{tab:predictive_state_eval}, PSM-WM achieves the
lowest PCG and RHG while obtaining the strongest future latent retrieval
performance. The reduced PCG indicates that predictive sufficiency guided
state construction preserves more information required for future prediction. The low RHG suggests that the learned state leaves less residual predictive information in the observed history under the evaluated probe setting. The retrieval results further show that the learned states capture latent structures aligned with future interaction semantics rather than only optimizing prediction outputs. These analyses provide empirical evidence, not a formal guarantee, that PSM-WM learns
future-oriented latent states with stronger predictive properties.

\begin{table*}[htbp]
\centering
\small
\setlength{\tabcolsep}{1.8pt} 
\renewcommand{\arraystretch}{1.05}
\caption{Evaluation of predictive state properties. PCG and RHG measure the remaining predictive gap and residual historical information, respectively. Future latent retrieval evaluates alignment with future semantics. Lower PCG/RHG and higher retrieval indicate stronger predictive sufficiency.}
\begin{tabular}{lccc|ccc|cc}
\toprule&
\multicolumn{3}{c|}{\textbf{Predictive Closure Gap} ($\downarrow$)}
&
\multicolumn{3}{c|}{\textbf{Residual History Test} (MELD)}
&
\multicolumn{2}{c}{\textbf{Future Latent Retrieval} ($\uparrow$)}
\\
\textbf{Method}&Hi-EF&IEMOCAP&MELD&State Only&State+History&RHG ($\downarrow$)&Recall@1&MRR\\
\midrule
\midrule
MISA~\cite{misa}
&8.72&1.84&5.46
&65.12&67.47&2.35
&32.80&0.440
\\
CPC~\cite{CPC}
&6.53&1.39&-
&-&-&-  
&40.51&0.543
\\
BYOL~\cite{BYOL}
&6.18&1.33&3.46
&67.12&68.86&1.74
&41.73&0.563
\\

RSSM~\cite{RSSM}

&5.12&1.06&1.37
&69.21&70.43&1.22
&43.82&0.591
\\

LPWM~\cite{LPWM}
&4.18&0.78&1.16
&69.42&70.24&0.82
&45.36&0.624
\\
HAUWM~\cite{HAUWM}
&4.02&0.64&1.03
&69.55&70.31&0.76
&45.72&0.636
\\
ModularAgent~\cite{ModularAgent}
&3.86&0.57&0.96
&69.62&70.28&0.66
&46.15&0.632
\\
\midrule
\midrule
\textbf{PSM-WM (Ours)}&\textbf{2.30}&\textbf{0.41}&\textbf{0.61}&
\textbf{69.97}&\textbf{70.58}&\textbf{0.61}&
\textbf{47.81}&\textbf{0.656}\\
\bottomrule
\end{tabular}
\label{tab:predictive_state_eval}
\end{table*}

\begin{figure}[htbp]
    \centering
    \includegraphics[width=1\linewidth]{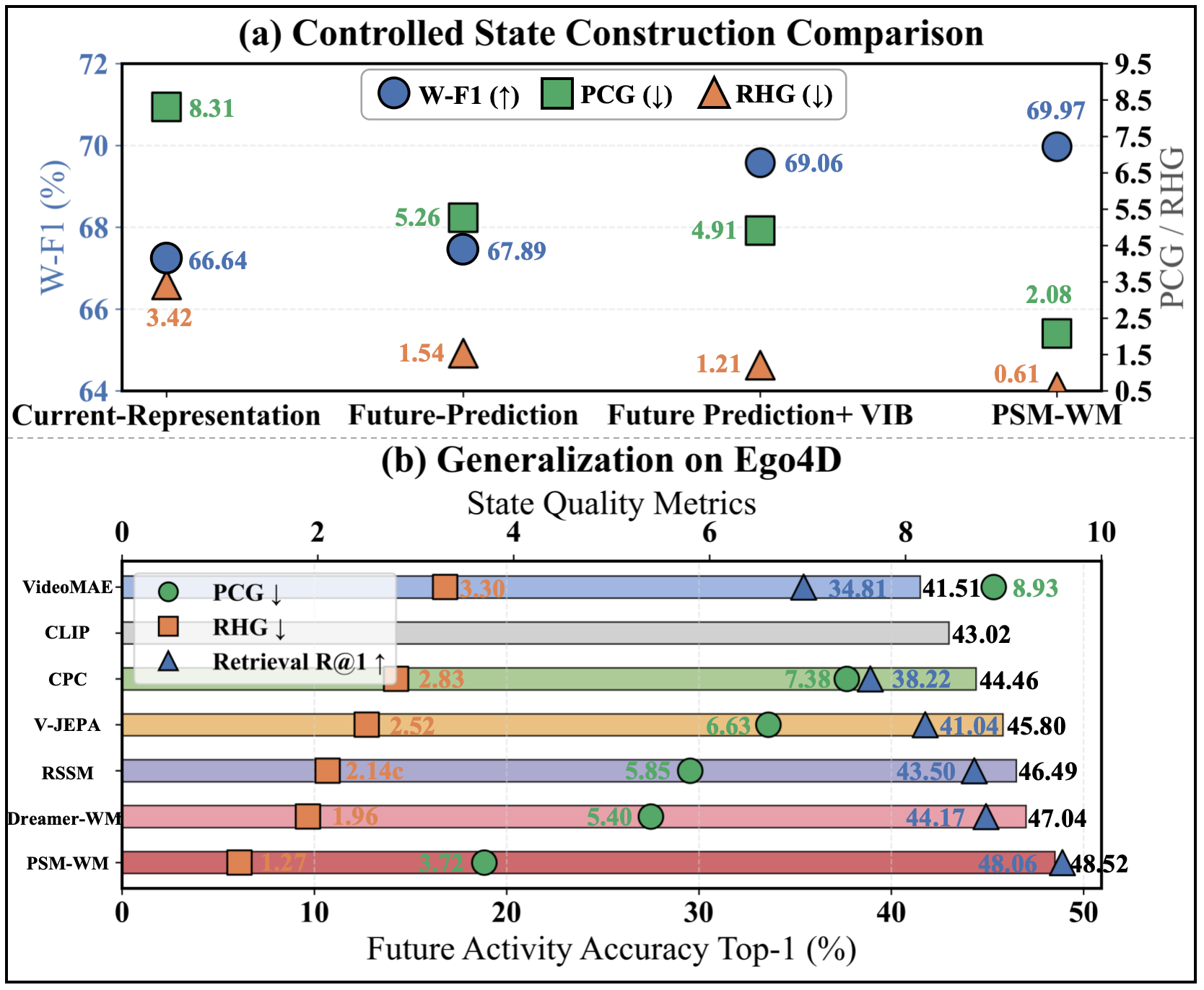}
   \caption{(a) Predictive sufficiency guided state construction improves future modeling on MELD under controlled settings with identical architectures and optimization budgets. (b) PSM-WM generalizes to Ego4D by preserving future-relevant information beyond social interaction scenarios.}
    \label{fig:5+9}
\end{figure}


\subsection{Does Predictive State Construction Matter?}

We hypothesize that latent state quality is determined by the information
preserved during state construction rather than solely by transition
modeling. To verify this, we perform a controlled comparison where all
components except state construction are fixed, including the multimodal
encoder, latent dimension, transition model, event representation, future
target, and optimization budget. We compare four state construction strategies: Current Representation, Future Prediction State, Future Prediction with VIB, and PSM-WM. Current Representation optimizes states for present semantic prediction, while Future Prediction State learns future-aware representations without explicit information constraints. Future Prediction with VIB further introduces generic compression through a variational information bottleneck~\cite{alemi2017deep}. As shown in Fig.~\ref{fig:5+9}(a), future prediction improves over current
representation learning, and VIB provides additional compression benefits. However, PSM-WM achieves the lowest PCG and RHG, indicating that predictive state construction is more effective than generic future prediction or compression.


\subsection{RQ3: Predictive State Evolution Analysis}

Predictive states determine what information is preserved, while world
models require transition mechanisms to model how such states evolve.
Therefore, we analyze the contribution of state construction and latent
dynamics from three aspects: predictive state construction, 
event-conditioned stochastic transitions, and long-horizon rollout stability. \textbf{Effect of Predictive State Construction.}
We first isolate the impact of state construction by fixing the transition model and only replacing the latent representation. As shown in Fig.~\ref{fig:PSE-Analysis}(a), predictive states consistently outperform generic latent representations, demonstrating that preserving
future-relevant information provides a more effective basis for future
interaction modeling. \textbf{Effect of Event-conditioned Stochastic Transition.} Using the same predictive state representation, we further evaluate the transition mechanism. Event conditioning improves future prediction by incorporating interaction context, while stochastic transitions further enhance performance by capturing diverse evolution trajectories (Fig.~\ref{fig:PSE-Analysis}(b)). \textbf{Long-horizon Rollout Stability.} We finally evaluate recursive latent rollout under different prediction
horizons. PSM-WM maintains slower performance degradation as the rollout horizon increases, indicating that predictive states provide a more stable latent space for modeling future evolution (Fig.~\ref{fig:PSE-Analysis}(c)).

\begin{figure*}[ht]
    \centering
    \includegraphics[width=1\linewidth]{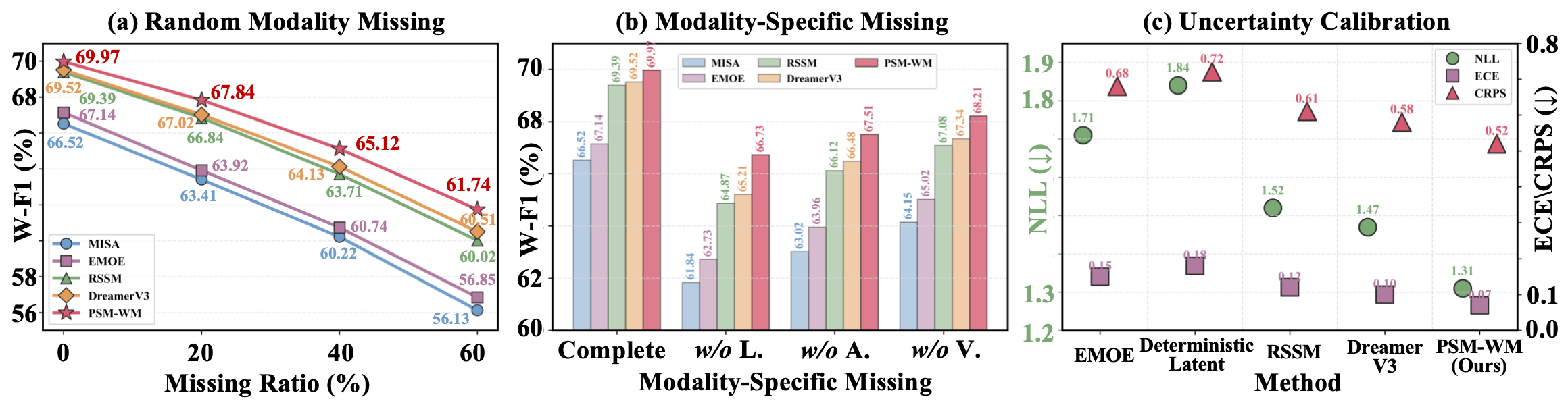}
    \caption{Robustness and uncertainty calibration analysis of PSM-WM.
(a) Future prediction performance on MELD under varying random modality missing ratios. (b) Robustness on MELD under modality-specific missing conditions. (c) Calibration quality of predictive uncertainty using NLL, ECE, and CRPS.}
    \label{fig:Robuness+Uncertainty}
\end{figure*}

\subsection{RQ4: Robustness under Incomplete Observations and Uncertainty}

Multimodal interactions often suffer from missing observations and uncertain
future dynamics. Therefore, we evaluate whether PSM-WM maintains reliable
future prediction under incomplete evidence and whether its uncertainty
estimates are properly calibrated. \textbf{Robustness under Incomplete Observations.}
We evaluate robustness under both random modality missing ratios and
modality-specific missing scenarios. All methods degrade with increasing
missing observations, while PSM-WM exhibits slower performance degradation (Fig.~\ref{fig:Robuness+Uncertainty}(a)). The modality-specific results in Fig.~\ref{fig:Robuness+Uncertainty}(b)
further show consistent improvements across different missing patterns.
These results indicate that probabilistic belief inference helps preserve
future-relevant information when multimodal evidence is incomplete. \textbf{Uncertainty Calibration Analysis.}
We further evaluate whether the estimated uncertainty reflects prediction
reliability using negative log-likelihood (NLL), expected calibration error (ECE), and continuous ranked probability score (CRPS). PSM-WM achieves better calibration than deterministic representations and stochastic latent dynamics models (Fig.~\ref{fig:Robuness+Uncertainty}(c)).

\begin{table}[t]
\centering
\setlength{\tabcolsep}{1.8pt} 
\renewcommand{\arraystretch}{1.05}
\caption{Ablation study of PSM-WM.}
\label{tab:ablation}
\begin{tabular}{lcc}
\toprule
\textbf{Variant}& \textbf{MELD W-F1}& \textbf{Ego4D Acc.}\\
\midrule
\textbf{PSM-WM}&\textbf{69.97}&\textbf{48.5}\\
\midrule
\textit{w/o} Predictive Sufficiency&68.76&46.9\\
\textit{w/o} Belief State&69.12&47.4\\
\textit{w/o} Event Conditioning&69.31&47.7\\
\textit{w/o} Stochastic Transition&69.42&47.8\\
\textit{w/o} Rollout Consistency&69.28&47.3\\
\bottomrule
\end{tabular}
\end{table}

\subsection{Ablation Studies}

We conduct ablation studies to examine the contribution of each component
in PSM-WM, including predictive sufficiency guided state construction,
probabilistic belief inference, event-conditioned transitions, stochastic
dynamics, and rollout consistency. All variants are trained and evaluated
under the same protocol as the full model. We additionally evaluate the
ablated models on Ego4D to examine the generality of each component beyond social interaction scenarios. Removing any component consistently
degrades performance across both benchmarks (Table~\ref{tab:ablation}). The largest degradation is observed when removing the predictive sufficiency objective, highlighting the importance of future-oriented state construction for learning predictive latent representations. Removing probabilistic belief inference also reduces performance, demonstrating the benefit of uncertainty-aware state estimation under incomplete multimodal observations. Event-conditioned transitions, stochastic dynamics, and rollout consistency further contribute to modeling context-dependent and long-horizon state evolution. These results show that PSM-WM relies on both state construction and latent dynamics modeling.

\subsection{Generalization to Egocentric Multimodal World Modeling}

To evaluate whether predictive sufficiency extends beyond social interaction, we further test PSM-WM on Ego4D. Compared with conversational scenarios, Ego4D involves different temporal dynamics and partial observability, providing a broader evaluation of future-oriented state construction. We compare PSM-WM with current-state representation methods (VideoMAE~\cite{tong2022videomae} and CLIP-based~\cite{radford2021learning} temporal encoders), predictive representation methods (CPC and V-JEPA), and latent world models (RSSM-based and Dreamer-style models). Future Evolution Modeling Beyond Social Interaction. Fig.~\ref{fig:5+9}(b) reports future prediction performance and state-centric evaluations on Ego4D. PSM-WM achieves stronger future prediction while obtaining lower predictive closure gap and residual history gain, together with improved future latent retrieval. These results indicate that predictive sufficiency guided state construction preserves future-relevant information in a different domain.





\begin{figure}[htbp]
    \centering
    \includegraphics[width=1\linewidth]{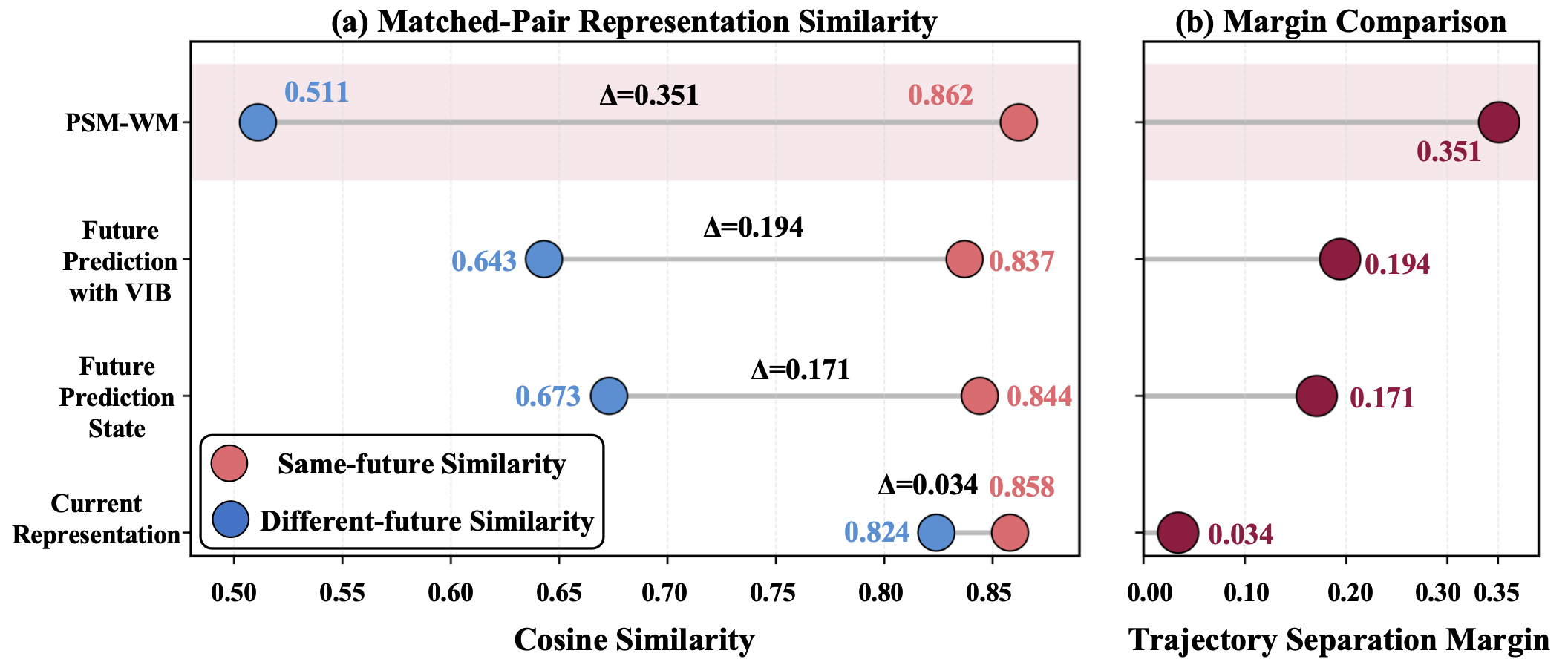}
   \caption{Representation similarity for matched sample pairs with identical current semantics. Each line connects the different-future and same-future similarities of one method, and the line length indicates the trajectory separation margin. PSM-WM achieves the largest separation while maintaining the highest same-future similarity.}
    \label{fig:future_trajectory_discrimination}
\end{figure}

\subsection{Current Semantic Similarity Does Not Guarantee Future Similarity}
\label{sec:future_trajectory_discrimination}

To examine whether learned states distinguish future evolution beyond
current semantics, we construct matched sample pairs with identical
current labels but either matching or different future label sequences
over the prediction horizon. We measure cosine similarity between
frozen representations and define the trajectory separation margin as
$\mathcal{M}_{\mathrm{traj}}=
\operatorname{Sim}_{\mathrm{same}}-
\operatorname{Sim}_{\mathrm{diff}}$.

As shown in Fig.~\ref{fig:future_trajectory_discrimination},
the margin increases from 0.034 for Current Representation to
0.171 for Future Prediction State and 0.194 for Future Prediction
with VIB. PSM-WM achieves the largest margin of 0.351, with the
highest same-future similarity (0.862) and lowest different-future
similarity (0.511). This indicates that predictive sufficiency
guided state construction preserves representation consistency
for matching future trajectories while better distinguishing
different future evolutions, beyond generic future prediction
and information compression.

\section{Conclusion}

This work investigates how multimodal latent states should preserve
historical information to support future interaction modeling.
To address the limitations of current-state representation learning
and direct future prediction, we propose PSM-WM, which constructs
latent states guided by predictive sufficiency and models their
evolution through probabilistic belief inference and
event-conditioned stochastic transitions. Experiments across multiple benchmarks demonstrate improved future
interaction modeling. Further analyses validate the predictive
properties of learned states, their ability to distinguish different
future trajectories, and their robustness to incomplete observations.
Controlled comparisons and ablations further establish the
complementary contributions of predictive state construction and
latent dynamics modeling.

Future work will explore longer-horizon and action-conditioned
prediction, improve uncertainty estimation under distribution shifts,
and extend PSM-WM to real-world interactive environments.



\newpage


{
    \small
    \bibliographystyle{ieeenat_fullname}
    \bibliography{main}

@String(CVPR= {IEEE Conf. Comput. Vis. Pattern Recog.})

@String(ICASSP=	{ICASSP})

@String(AAAI = {AAAI})

@String(CVPR  = {CVPR})

@inproceedings{Aoept,
  title={{AOEPT}: Breaking the Implicit Modality-Reduction Bottleneck in Modality-Missing Prompt Tuning},
  author={Lang, Jian and Hong, Rongpei and Zhong, Ting and Zhou, Fan},
  booktitle={Proceedings of the 43rd International Conference on Machine Learning},
  series={Proceedings of Machine Learning Research},
  volume={306},
  pages={62583--62602},
  publisher={PMLR},
  year={2026}
}

@inproceedings{MULT,
  title={Multimodal transformer for unaligned multimodal language sequences},
  author={Tsai, Yao-Hung Hubert and Bai, Shaojie and Liang, Paul Pu and Kolter, J Zico and Morency, Louis-Philippe and Salakhutdinov, Ruslan},
  booktitle={Proceedings of the 57th annual meeting of the association for computational linguistics},
  pages={6558--6569},
  year={2019}
}

@inproceedings{misa,
  title={{MISA: Modality-Invariant and -Specific Representations for Multimodal Sentiment Analysis}},
  author={Hazarika, Devamanyu and Zimmermann, Roger and Poria, Soujanya},
  booktitle={Proceedings of the 28th ACM International Conference on Multimedia (ACM MM)},
  pages={1122--1131},
  year={2020},
  publisher = {Association for Computing Machinery},
  address = {New York, NY, USA}
}

@inproceedings{EMOE,
  title={{EMOE}: Modality-Specific Enhanced Dynamic Emotion Experts},
  author={Fang, Yiyang and Huang, Wenke and Wan, Guancheng and Su, Kehua and Ye, Mang},
  booktitle={Proceedings of the IEEE/CVF Conference on Computer Vision and Pattern Recognition (CVPR)},
  pages={14314--14324},
  year={2025},
  publisher = {IEEE/CVF}
}

@inproceedings{DLF,
  title={{DLF}: Disentangled-language-focused multimodal sentiment analysis},
  author={Wang, Pan and Zhou, Qiang and Wu, Yawen and Chen, Tianlong and Hu, Jingtong},
  booktitle={Proceedings of the AAAI Conference on Artificial Intelligence (AAAI)},
  pages={21180--21188},
  year={2025},
  publisher = {AAAI Press}
}

@InProceedings{CLCR,
    author    = {Meng, Chunlei and Huang, Guanhong and Fu, Rong and Jian, Runmin and Gan, Zhongxue and Ouyang, Chun},
    title     = {{CLCR}: Cross-Level Semantic Collaborative Representation for Multimodal Learning},
    booktitle = {Proceedings of the IEEE/CVF Conference on Computer Vision and Pattern Recognition (CVPR)},
    year      = {2026},
    pages     = {1606-1615},
    publisher = {IEEE/CVF}
}

@InProceedings{TSD,
    author    = {Meng, Chunlei and Luo, Jiabin and Yan, Zhenglin and Yu, Zhenyu and Fu, Rong and Gan, Zhongxue and Ouyang, Chun},
    title     = {Tri-Subspaces Disentanglement for Multimodal Sentiment Analysis},
    booktitle = {Proceedings of the IEEE/CVF Conference on Computer Vision and Pattern Recognition (CVPR)},
    year      = {2026},
    pages     = {8791-8800},
    publisher = {IEEE/CVF}
}

@inproceedings{TSDA,
  author={Meng, Chunlei and Zhou, Ziyang and He, Lucas and Du, Xiaojing and Ouyang, Chun and Gan, Zhongxue},
  booktitle={ICASSP 2026 - 2026 IEEE International Conference on Acoustics, Speech and Signal Processing (ICASSP)}, 
  title={Temporal-Spatial Decouple Before Act: Disentangled Representation Learning for Multimodal Sentiment Analysis}, 
  pages={11917-11921},
  publisher = {IEEE},
  year={2026}
}

@misc{MRCF,
      title={Rethinking Modality Reliability in Multimodal Sentiment Analysis with Incomplete Observations}, 
      author={Chunlei Meng and Jacqueline J. Pang and Pengbin Feng and Zhenyu Yu and Chun Ouyang and Zhongxue Gan},
      year={2026},
      eprint={2608.03611},
      archivePrefix={arXiv}
}

@InProceedings{GCL,
  title =   {Group Cognition Learning: Making Everything Better Through Governed Two-Stage Agents Collaboration},
  author =       {Meng, Chunlei and Feng, Pengbin and Fu, Rong and Lee, Hoi Leong and Du, Xiaojing and Kang, Zhaolu and Zhang, Zeyu and Zhou, Weilin and Ouyang, Chun and Gan, Zhongxue},
  booktitle =   {Proceedings of the 43rd International Conference on Machine Learning},
  pages =   {87822--87834},
  year =   {2026},
  volume =   {306},
  series =   {Proceedings of Machine Learning Research},
  publisher =    {PMLR},
  url =   {https://proceedings.mlr.press/v306/meng26a.html}
}

@inproceedings{DBR,
  author = {Meng, Chunlei and Luo, Jiabin and Feng, Pengbin and Yan, Zhenglin and Hu, Chengyin and Gan, Zhongxue and Ouyang, Chun},
  title = {Mitigating Shared-Private Branch Imbalance via Dual-Branch Rebalancing for Multimodal Sentiment Analysis},
  publisher = {Association for Computing Machinery},
  booktitle = {Proceedings of the 34th ACM International Conference on Multimedia},
  pages = {7690–7698},
  numpages = {9},
  year = {2026}
}

@inproceedings{EMART,
    title = {Emotion-Wheel-Guided Audio-Referred Text Representation for Multimodal Emotion Recognition in Conversation},
    author = {Seong, Eunseon and Lee, Harim and Kim, Dahye and Kim, Changhyun and Chae, Dong-Kyu},
    booktitle = {Proceedings of the 64th Annual Meeting of the {A}ssociation for {C}omputational {L}inguistics (Volume 1: Long Papers)},
    year = {2026},
    publisher = {Association for Computational Linguistics},
    pages = {40392--40403}
}

@inproceedings{HyperEF,
  title={Beyond Missing Modalities: Hypergraph Conditioned Diffusion for Uncertainty-Aware Multimodal Emotion Recognition},
  author={Qiu, Xihang and Fang, Yuhao and Zhou, Qing and Zhai, Bin and Hong, Jialong and Zhang, Wanpeng and Lu, Yao and Zhang, Ye and Li, Chun},
  booktitle={Proceedings of the IEEE/CVF Conference on Computer Vision and Pattern Recognition},
  pages={22953--22963},
  year={2026}
}

@inproceedings{Hi-EF,
  title={{Hi-EF}: Benchmarking Emotion Forecasting in Human-interaction},
  author={Wang, Haoran and Mai, Xinji and Tao, Zeng and Lin, Junxiong and Tong, Xuan and Pan, Ivy and Yan, Shaoqi and Wang, Yan and Gao, Shuyong},
  booktitle={Proceedings of the AAAI Conference on Artificial Intelligence},
  pages={2110--2118},
  year={2026}
}

@inproceedings{SWM,
title={Building Social World Models with Large Language Models},
author={Haofei Yu and Yining Zhao and Guanyu Lin and Jiaxuan You},
booktitle={Proceedings of the 43rd International Conference on Machine Learning},
series={Proceedings of Machine Learning Research},
volume={306},
pages={151719--151736},
publisher={PMLR},
year={2026},
url={https://openreview.net/forum?id=Wxqr2fkZmX}
}

@inproceedings{HAUWM,
  title={Learning to Be Uncertain: Pre-training World Models with Horizon-Calibrated Uncertainty},
  author={Wan, Shenghua and Gan, Le and Zhan, De-Chuan},
  booktitle={International Conference on Learning Representations},
  volume={2026},
  pages={106107--106127},
  year={2026}
}

@article{IEMOCAP,
  title={{IEMOCAP}: Interactive emotional dyadic motion capture database},
  author={Busso, Carlos and Bulut, Murtaza and Lee, Chi-Chun and Kazemzadeh, Abe and Mower, Emily and Kim, Samuel and Chang, Jeannette N and Lee, Sungbok and Narayanan, Shrikanth S},
  journal={Language Resources and Evaluation},
  volume={42},
  number={4},
  pages={335--359},
  year={2008},
  publisher={Springer}
}

@inproceedings{meld,
  title={{MELD}: A Multimodal Multi-Party Dataset for Emotion Recognition in Conversations},
  author={Poria, Soujanya and Hazarika, Devamanyu and Majumder, Navonil and Naik, Gautam and Cambria, Erik and Mihalcea, Rada},
  booktitle={Proceedings of the 57th annual meeting of the association for computational linguistics},
  pages={527--536},
  year={2019}
}

@misc{lecun2022path,
  title={A Path Towards Autonomous Machine Intelligence},
  author={LeCun, Yann},
  howpublished={OpenReview},
  year={2022},
  note={Version 0.9.2, June 27, 2022},
  url={https://openreview.net/forum?id=BZ5a1r-kVsf}
}

@article{CPC,
  title={Representation learning with contrastive predictive coding},
  author={Oord, Aaron van den and Li, Yazhe and Vinyals, Oriol},
  journal={arXiv preprint arXiv:1807.03748},
  year={2018}
}

@article{BYOL,
  title={Bootstrap Your Own Latent - A New Approach to Self-Supervised Learning},
  author={Grill, Jean-Bastien and Strub, Florian and Altch{\'e}, Florent and Tallec, Corentin and Richemond, Pierre and Buchatskaya, Elena and Doersch, Carl and Avila Pires, Bernardo and Guo, Zhaohan and Gheshlaghi Azar, Mohammad and others},
  journal={Advances in neural information processing systems},
  volume={33},
  pages={21271--21284},
  year={2020}
}

@InProceedings{ModularAgent,
    author    = {Zhan, Yu-Wei and Wang, Xin and Mao, Pengzhe and Feng, Tongtong and Wang, Ren and Zhu, Wenwu},
    title     = {{ModularAgent}: A Task-Aware Modular Framework for Joint Optimization of Multimodal Large Language Models and World Models},
    booktitle = {Proceedings of the IEEE/CVF Conference on Computer Vision and Pattern Recognition (CVPR)},
    year      = {2026},
    pages     = {8087-8096}
}

@inproceedings{LPWM,
 author = {Daniel, Tal and Qi, Carl and Haramati, Dan and Zadeh, Amir and Li, Chuan and Tamar, Aviv and Pathak, Deepak and Held, David},
 booktitle = {International Conference on Learning Representations},
 pages = {127719--127767},
 title = {Latent Particle World Models: Self-supervised Object-centric Stochastic Dynamics Modeling},
 year = {2026}
}

@inproceedings{RSSM,
  author       = {Danijar Hafner and Timothy P. Lillicrap and Ian Fischer and Ruben Villegas and David Ha and Honglak Lee and James Davidson},
  title        = {Learning Latent Dynamics for Planning from Pixels},
  booktitle    = {Proceedings of the 36th International Conference on Machine Learning,
                  {ICML} 2019},
  pages        = {2555--2565},
  year         = {2019}
}

@article{V-JEPA2,
      title={{V-JEPA} 2: Self-Supervised Video Models Enable Understanding, Prediction and Planning},
      journal={arXiv preprint arXiv:2506.09985},
      author={Mido Assran and Adrien Bardes and David Fan and Quentin Garrido and Russell Howes and Komeili, Mojtaba and Matthew Muckley and Ammar Rizvi and Claire Roberts and Koustuv Sinha and Artem Zholus and Sergio Arnaud and Abha Gejji and Ada Martin and Francois Robert Hogan and Daniel Dugas and Piotr Bojanowski and Vasil Khalidov and Patrick Labatut and Francisco Massa and Marc Szafraniec and Kapil Krishnakumar and Yong Li and Xiaodong Ma and Sarath Chandar and Franziska Meier and Yann LeCun and Michael Rabbat and Nicolas Ballas},
      year={2025},
      eprint={2506.09985},
      archivePrefix={arXiv},
      primaryClass={cs.AI},
      url={https://arxiv.org/abs/2506.09985}, 
}

@article{DreamerV3,
  title={Mastering diverse control tasks through world models},
  author={Hafner, Danijar and Pasukonis, Jurgis and Ba, Jimmy and Lillicrap, Timothy},
  journal={Nature},
  volume={640},
  number={8059},
  pages={647--653},
  year={2025},
  publisher={Nature Publishing Group}
}

@inproceedings{AHRA,
    title = {Adaptive Hierarchical Representation Alliance for Multimodal Learning},
    author = {Meng, Chunlei and Feng, Pengbin and Pang, Jacqueline J.  and Liao, Chih-Ting and Fu, Rong  and Kang, Zhaolu  and Gan, Zhongxue  and Ouyang, Chun},
    booktitle = {Findings of the {A}ssociation for {C}omputational {L}inguistics: {EMNLP} 2026},
    year = {2026},
    publisher = "Association for Computational Linguistics",
    pages = {21423--21439}
}

@InProceedings{Ego4D,
    author    = {Grauman, Kristen and Westbury, Andrew and Byrne, Eugene and Chavis, Zachary and Furnari, Antonino and Girdhar, Rohit and Hamburger, Jackson and Jiang, Hao and Liu, Miao and Liu, Xingyu and Martin, Miguel and Nagarajan, Tushar and Radosavovic, Ilija and Ramakrishnan, Santhosh Kumar and Ryan, Fiona and Sharma, Jayant and Wray, Michael and Xu, Mengmeng and Xu, Eric Zhongcong and Zhao, Chen and Bansal, Siddhant and Batra, Dhruv and Cartillier, Vincent and Crane, Sean and Do, Tien and Doulaty, Morrie and Erapalli, Akshay and Feichtenhofer, Christoph and Fragomeni, Adriano and Fu, Qichen and Gebreselasie, Abrham and Gonz\'alez, Cristina and Hillis, James and Huang, Xuhua and Huang, Yifei and Jia, Wenqi and Khoo, Weslie and Kol\'a\v{r}, J\'achym and Kottur, Satwik and Kumar, Anurag and Landini, Federico and Li, Chao and Li, Yanghao and Li, Zhenqiang and Mangalam, Karttikeya and Modhugu, Raghava and Munro, Jonathan and Murrell, Tullie and Nishiyasu, Takumi and Price, Will and Ruiz, Paola and Ramazanova, Merey and Sari, Leda and Somasundaram, Kiran and Southerland, Audrey and Sugano, Yusuke and Tao, Ruijie and Vo, Minh and Wang, Yuchen and Wu, Xindi and Yagi, Takuma and Zhao, Ziwei and Zhu, Yunyi and Arbel\'aez, Pablo and Crandall, David and Damen, Dima and Farinella, Giovanni Maria and Fuegen, Christian and Ghanem, Bernard and Ithapu, Vamsi Krishna and Jawahar, C. V. and Joo, Hanbyul and Kitani, Kris and Li, Haizhou and Newcombe, Richard and Oliva, Aude and Park, Hyun Soo and Rehg, James M. and Sato, Yoichi and Shi, Jianbo and Shou, Mike Zheng and Torralba, Antonio and Torresani, Lorenzo and Yan, Mingfei and Malik, Jitendra},
    title     = {{Ego4D}: Around the World in 3,000 Hours of Egocentric Video},
    booktitle = {Proceedings of the IEEE/CVF Conference on Computer Vision and Pattern Recognition (CVPR)},
    month     = {June},
    year      = {2022},
    pages     = {18995-19012}
}

@inproceedings{alemi2017deep,
  title     = {Deep Variational Information Bottleneck},
  author    = {Alemi, Alexander A. and Fischer, Ian and Dillon, Joshua V. and Murphy, Kevin},
  booktitle = {International Conference on Learning Representations},
  year      = {2017}
}

@inproceedings{tong2022videomae,
  title     = {{VideoMAE}: Masked Autoencoders are Data-Efficient Learners for Self-Supervised Video Pre-Training},
  author    = {Tong, Zhan and Song, Yibing and Wang, Jue and Wang, Limin},
  booktitle = {Advances in Neural Information Processing Systems},
  volume    = {35},
  pages     = {10078--10093},
  year      = {2022}
}

@inproceedings{radford2021learning,
  title     = {Learning Transferable Visual Models From Natural Language Supervision},
  author    = {Radford, Alec and Kim, Jong Wook and Hallacy, Chris and Ramesh, Aditya and Goh, Gabriel and Agarwal, Sandhini and Sastry, Girish and Askell, Amanda and Mishkin, Pamela and Clark, Jack and Krueger, Gretchen and Sutskever, Ilya},
  booktitle = {Proceedings of the 38th International Conference on Machine Learning},
  series    = {Proceedings of Machine Learning Research},
  volume    = {139},
  pages     = {8748--8763},
  publisher = {PMLR},
  year      = {2021}
}
}

\end{document}